\input{aipcheck}

\documentclass[
    ,final            
  ]
  {aipproc}

\layoutstyle{6x9}

\def\massG{~GeV}
\def\invfb{fb$^{-1}$}

\newcommand {\gapprox}
   {\raisebox{-0.7ex}{$\stackrel {\textstyle>}{\sim}$}}
\newcommand {\lapprox}
   {\raisebox{-0.7ex}{$\stackrel {\textstyle<}{\sim}$}}

\begin{document}

\title{Discovery Potential of $R$-hadrons with the ATLAS Detector at the LHC}


\author{Philippe Mermod, for the ATLAS Collaboration}{
  address={Department of Physics, Stockholm University, SE-10691 Stockholm, Sweden}
  ,address={Subdepartment of Particle Physics, University of Oxford, Oxford OX1 3NP, UK}
}

\begin{abstract}
$R$-hadrons are predicted in a range of supersymmetric scenarios including split-supersymmetry and gauge-mediated 
supersymmetry breaking. In this paper, the discovery potential of the ATLAS experiment for gluino and stop-based 
$R$-hadrons is outlined. A range of final state observables such as high transverse momentum muon-like objects 
and information on ionization energy loss is used. It is shown that ATLAS would be able to discover such 
particles at comparatively modest amounts of luminosity (1\invfb) for masses up to 1 TeV. 
\end{abstract}

\maketitle



%
%
%
%
%

\section{Introduction}
The presence (or absence) of massive exotic stable hadrons will be
an important observable in the search for and quantification of any
new physics processes seen at the LHC. Stable exotic coloured
particles are predicted in a range of SUSY scenarios (see, for
example, Ref.~\cite{Fairbairn}).
Such particles could be copiously produced at the LHC and
sensitivity to particle masses substantially beyond those excluded
by earlier collider searches ($\lapprox 200$ GeV~\cite{Fairbairn}) 
could be achieved at ATLAS even with
rather modest amounts of integrated luminosity ($\sim 1$fb$^{-1}$)~\cite{CSC08}.
This paper outlines a strategy for the detection of exotic
massive, long-lived hadrons (so-called $R$-hadrons) formed from
either stable gluinos or stops ($R_{\tilde{g}}$ and
$R_{\tilde{t}}$-hadrons, respectively). The $R_{\tilde{g}}$-hadrons
($R_{\tilde{t}}$-hadrons) are considered in the context of a
Split-SUSY (stop NLSP/gravitino LSP) scenario \cite{splitSUSY}. Although this work is
performed in the framework of SUSY, the techniques presented here
may be used in generic searches for stable heavy exotic hadrons. 

\section{Event generation}
\begin{figure}[htbp]
    \includegraphics[width=0.75\linewidth]{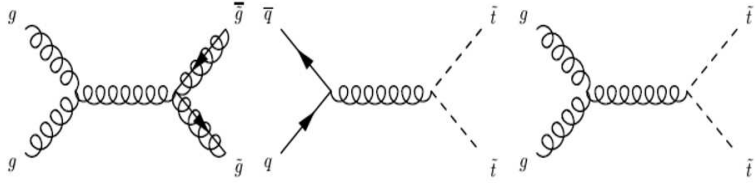}
  \caption{Selection of leading-order processes illustrating the production of
  gluino and stop particles.
  \label{fig:feynman}}
\end{figure}

The leading-order event generator {\sc PYTHIA}~\cite{PYTHIA} was used to produce samples of
pair-produced gluino and stop-antistop events for a range of gluino
and stop masses between 300 and 2000 GeV. The number of expected pair-production 
events for 14 TeV proton-proton collisions with $1\mathrm{fb}^{-1}$ of integrated 
luminosity is $2.7\times10^5$ and $7.8\times10^{3}$ for 300 GeV gluino pairs and stop pairs, 
respectively; for a mass of 1000 GeV, it is $138$ and $6$, respectively. 
Production mechanisms of stops and gluinos are illustrated in
Figure~\ref{fig:feynman}, which shows leading-order Feynman diagrams. 
The cross section depends principally on the mass of the heavy object under study and not
other free SUSY parameters.

To complement the signal samples, various background samples were
used, each corresponding to an integrated luminosity of at least
 $\sim 1$~\invfb. 
As the simulated trigger used for this analysis 
requires a hard muon-like track, only events which could give rise
to a high $p_T$-muon ($p_T > 150$~GeV) were simulated. The following
processes were considered : leading-order 2-to-2 QCD processes, which
include all quark flavours except top ; backgrounds 
arising from diboson and single boson production, denoted electroweak ; 
and a sample of $t\bar{t}$ pair-production events, termed
top.

\section{Simulation of $R$-hadron scattering in matter}
\label{matter} 

A model of $R$-hadron scattering~\cite{Mackeprang} 
implemented in Geant4~\cite{Allison:2006ve} is used in this work.
This is an update of earlier work \cite{Kraan} in which the geometric 
cross-section and phase space arguments are used to predict the different 2-to-2 and 2-to-3
reactions. Other approaches to modelling $R$-hadron scattering have
been proposed, based on Regge phenomenology~\cite{Dave}. These
yield predictions of energy loss and scattering cross-sections which
are qualitatively similar to those given by the model used here~\cite{Dave2}.

The typical energy loss per interaction is predicted to
be low (around several~GeV~\cite{Mackeprang}). 
This implies that the fraction of $R$-hadrons which would be stopped during their 
traversal of the detector is negligible\footnote{Although it does not form a part
of this work, the possibility of observing the decay of stopped $R$-hadrons 
offers a promising and complementary means of 
searching for $R$-hadrons at the LHC~\cite{Arvanitaki,D0stopped}.}.

Another feature is the possibility of charge and baryon number exchange. 
Following repeated scattering $R_{\tilde{g}}$-hadrons and
$R_{\tilde{t}}$-hadrons not containing an anti-stop should enter the
muon system predominantly as baryons. This is due to the occurence
of meson-to-baryon conversion processes for which the inverse
reaction is suppressed~\cite{Kraan}. Anti-baryons 
would be expected to quickly annihilate in matter and
$R_{\tilde{t}}$-hadrons containing anti-stops would thus largely
remain as mesons. In the model used here, charged $R$-baryon states 
are stable and thus are expected to leave tracks in the muon system  
(although this assumption is now disputed in the case of the $R_{\tilde{g}}$-baryons \cite{Dave2}).    
Also, a substantial rate of events is expected in 
which a $R$-hadron appears to possess different values of the electric charge  
in the inner detector and muon system. 
While such topologies represent a challenge for track reconstruction 
software, they also provide observables useful for the discovery and 
characterisation of $R$-hadrons (see next section).

\section{Event selection}
\label{evsel}

The selected level 1 trigger is the ${mu6}$ trigger~\cite{CSC08}, 
which is sensitive to the `classic' stable massive particle signature 
of a high transverse momentum muon-like track.
Here, we consider events in which a $R$-hadron
track in the muon system must be associated with the correct bunch
crossing. This leads to a rapid fall in efficiency for 
$\beta \lapprox 0.6$. 
After including requirements that the event filter is passed and the 
muon-like track is well-reconstructed, the overall efficiency 
is around 15\% for $R_{\tilde{g}}$-hadrons and around 25\% for 
$R_{\tilde{t}}$-hadrons, with little mass dependence. 
Ongoing work involve the 
development of triggers which do not rely on linked inner 
detector-muon chamber tracks~\cite{CSC08}, thus improving the overall efficiency by a 
factor of $2-3$ for $\beta \gapprox 0.6$.

Following the trigger selection, reconstructed final state
quantities were used to select $R$-hadron events and suppress
backgrounds, including (details are shown in Ref. \cite{CSC08}) : 
\begin{itemize}
\item the transverse momenta of muon-like tracks (the background events tend to be softer),  
\item the ratio of high and low threshold $HT/LT$ TRT hit
multiplicities (owing to their $\beta$ range the simulated $R$-hadron data peak at lower values of
$HT/LT$), 
\item the distance $R=({\Delta}{\eta}^2 + {\Delta}{\phi}^2)^{1/2}$ 
between a $R$-hadron track candidate and a jet 
with $p_T >100$~GeV (background muons are often associated or close 
to a hard jet while $R$ is larger in average for $R$-hadrons),  
\item the cosine of the angle between two $R$-hadron candidates 
which both leave hard tracks in either the inner detector or the muon system $\cos{\Delta\Phi}$ 
($R$-hadrons will be produced approximately back-to-back, unlike a number of background sources),
\item the variable $\frac{q_{ID}p_{T,ID}}{q_{\mu}p_{T\mu}}$, where $q_{ID}$ ,$q_{\mu}$,
$p_{T,ID}$, and $p_{T\mu}$ are the charge as reconstructed in the 
inner detector and muon system, and the reconstructed transverse 
momentum in the inner detector and muon system, respectively  
(as described in the previous section, the value of the $R$-hadron electric charge 
may be different in the inner detector and muon systems).  
\end{itemize}



A candidate $R$-hadron must satisfy that  
no hard muon-like track ($p_T > 250$~GeV) can come within a distance $R<0.36$ of a
hard jet and fulfill at least one of the conditions listed below. 
For consistency the same selection is applied both for $R_{\tilde{g}}$
and $R_{\tilde{t}}$-hadrons though criteria 3-4 are only relevant
for $R_{\tilde{g}}$-hadrons.

\begin{enumerate}
\item The event contains at least one hard muon track with no linked inner
detector track. A linked track is defined such that the distance
$R=({\Delta}{\eta}^2 + {\Delta}{\phi}^2)^{1/2}$ between the
measurements in the ID and muon systems is less than 0.1.
\item The event contains two hard back-to-back ($\cos{\Delta\Phi}<-0.85$) ID tracks with the TRT
hit distribution satisfying $HT/LT<0.05$. 
\item The event contains two hard back-to-back (as defined above) like-sign muon
tracks.
\item The event contains at least one hard muon track with a hard
matching ID track of opposite charge fulfilling the condition
$p_{T,ID} > 0.5 p_{T\mu}$.
\end{enumerate}
 

Table \ref{tab:finsel} shows the acceptance numbers and rates for
the various samples. It can be seen that for $R$-hadron masses below
$1$~TeV ATLAS opens up a discovery window with 
integrated luminosity of the order of $1$~fb$^{-1}$ at 14 TeV center-of-mass energy.
For masses above $1$~TeV the rate of signal events is small, and is comparable to the 
expected background rate, so discovery would be challenging even with larger data-sets. 

The event selection outlined above does not 
fully exploit the capabilities of the ATLAS detector : for instance, timing information 
was not used. Ongoing studies indicate that it is possible to measure the speed of the $R$-hadron 
candidates using time-of-flight information from the muon RPC and the Tile Calorimeter, with sufficient 
accuracy to enhance significantly the background rejection rate and, in the case of a discovery, 
to allow a measurement of the stable massive particle mass.  


\begin{table}[htbp]
  \centering
  \caption{Number of events selected for the given samples. Background
  samples not mentioned here are rejected by the selection.}
\bigskip
  \begin{tabular}{|lll|}
    \hline
    {\bf Sample} & {\bf Accepted events} & {\bf Rate (Events / \invfb)}
    \\
    \hline
    300 \massG{} gluino  & 235 & $6.44\times 10^3$ \\
    600 \massG{} gluino  & 551 & $2.70\times 10^2$ \\
    1000 \massG{} gluino & 774 & $10.7$ \\
    1300 \massG{} gluino & 732 & $1.20$ \\
    1600 \massG{} gluino & 685 & $0.147$\\
    2000 \massG{} gluino & 546 & $1.26\times 10^{-2}$\\
    \hline
  300 \massG{} stop  & 78 & $70.0$ \\
    600 \massG{} stop  & 134 & $3.9$ \\
    1000 \massG{} stop & 170 & $0.1$ \\
    \hline
    QCD & 2 & $0.9$\\
    electroweak ($Z\rightarrow \mu\mu$) & 1 & 0.8\\
    \hline
  \end{tabular}
  \label{tab:finsel}
\end{table}


\section{Conclusion}
Stable massive exotic hadrons ($R$-hadrons) are predicted in a
number of SUSY scenarios. By exploiting the signature of a hard
penetrating particle which may undergo charge exchange in the
calorimeter and seemingly does not fall within a jet, ATLAS will be
able to discover $R$-hadrons for masses below $1$~TeV with
relatively low amounts of integrated luminosity ($\sim 1$fb$^{-1}$).

\end{document}